\documentclass[10pt,journal,compsoc]{IEEEtran}

\ifCLASSOPTIONcompsoc
  \usepackage[nocompress]{cite}
\else
  \usepackage{cite}
\fi

\usepackage{amsmath}
\usepackage{amssymb}
\usepackage{booktabs}
\usepackage{graphicx}
\usepackage[varqu,scaled=0.88]{zi4}
\usepackage{listings}
\usepackage{microtype}
\usepackage{multirow}
\usepackage{pifont}
\usepackage[table]{xcolor}
\usepackage{tikz}
\usetikzlibrary{decorations.pathreplacing,positioning,tikzmark,calc,fit}
\usepackage[most]{tcolorbox}
\usepackage[hidelinks]{hyperref}
\hypersetup{
  pdftitle={SpecRL: Reinforcement Learning with Test-Based Completeness Rewards for Formal Specification Synthesis},
  pdfsubject={Formal specification synthesis for Dafny programs},
  pdfauthor={Zhechong Huang, Zhao Zhang, Zeyu Sun, Huifeng Sun, and Yingfei Xiong},
  pdfkeywords={formal verification, specification synthesis, reinforcement learning, large language models, Dafny}
}

\newtcolorbox{findingbox}[1][]{%
  colback=gray!10,
  colframe=black!70,
  fonttitle=\bfseries,
  boxrule=0.5pt,
  arc=2pt,
  left=6pt, right=6pt, top=4pt, bottom=4pt,
  title={#1},
}

\definecolor{codegreen}{RGB}{40,120,40}
\definecolor{codegray}{RGB}{100,100,100}
\definecolor{codeframe}{RGB}{214,219,226}
\definecolor{codekw}{RGB}{28,72,138}
\definecolor{specblue}{RGB}{0,80,180}
\definecolor{proofpurple}{RGB}{140,40,140}
\definecolor{codebg}{RGB}{250,251,252}

\lstdefinelanguage{Dafny}{
  morekeywords=[1]{method,returns,var,while,if,else,return,function,predicate,datatype,class,module,import,match,case,new,assert,assume,print,forall,exists,in},
  morekeywords=[2]{requires,ensures},
  morekeywords=[3]{invariant,decreases,modifies},
  morekeywords=[4]{int,nat,bool,string,real,char,seq,set,map,array,object,true,false},
  sensitive=true,
  morecomment=[l]{//},
  morecomment=[s]{/*}{*/},
  morestring=[b]",
}

\lstdefinestyle{dafny}{
  language=Dafny,
  basicstyle=\small\ttfamily,
  keywordstyle=[1]\color{codekw}\bfseries,
  keywordstyle=[2]\color{specblue}\bfseries,
  keywordstyle=[3]\color{proofpurple}\bfseries,
  keywordstyle=[4]\color{codegreen},
  commentstyle=\color{codegray},
  stringstyle=\color{codegreen},
  backgroundcolor=\color{codebg},
  numbers=none,
  numberstyle=\tiny\color{codegray},
  breaklines=true,
  breakatwhitespace=true,
  tabsize=2,
  showstringspaces=false,
  frame=single,
  framerule=0.4pt,
  rulecolor=\color{codeframe},
  framesep=2.5pt,
  framexleftmargin=1.5pt,
  framexrightmargin=1.5pt,
  framextopmargin=1pt,
  framexbottommargin=1pt,
  xleftmargin=0.25em,
  xrightmargin=0.25em,
  aboveskip=5pt,
  belowskip=5pt,
  columns=fixed,
  basewidth=0.54em,
  keepspaces=true,
  literate={ <= }{{\ensuremath{\mkern1mu\leq\mkern1mu}}}4
           { >= }{{\ensuremath{\mkern1mu\geq\mkern1mu}}}4
           { ==> }{{\ensuremath{\mkern1mu\Rightarrow\mkern1mu}}}5
           {<=}{{\ensuremath{\leq}}}2
           {>=}{{\ensuremath{\geq}}}2
           {==>}{{\ensuremath{\Rightarrow}}}3
           {!=}{{\ensuremath{\neq}}}2,
  escapeinside={(*@}{@*)},
}

\lstdefinestyle{prompt}{
  language={},
  basicstyle=\footnotesize\ttfamily,
  backgroundcolor=\color{codebg},
  commentstyle=\color{codegray},
  stringstyle=\color{codegreen},
  numbers=none,
  breaklines=true,
  breakatwhitespace=true,
  tabsize=2,
  showstringspaces=false,
  frame=single,
  framerule=0.4pt,
  rulecolor=\color{codeframe},
  framesep=3pt,
  framexleftmargin=2pt,
  framexrightmargin=2pt,
  framextopmargin=1pt,
  framexbottommargin=1pt,
  xleftmargin=0.35em,
  xrightmargin=0.35em,
  aboveskip=5pt,
  belowskip=5pt,
  columns=flexible,
  keepspaces=true,
}

\let\citep\cite
\let\citet\cite
\providecommand{\Description}[1]{}

\newcommand{\sem}[1]{\lbrack\!\lbrack #1 \rbrack\!\rbrack}
\newcommand{\parhead}[1]{\smallskip\noindent\textit{#1.}\ }

\begin{document}

\title{SpecRL: Reinforcement Learning with Test-Based Completeness Rewards for Formal Specification Synthesis}

\author{Zhechong~Huang,
        Zhao~Zhang,
        Zeyu~Sun,
        Huifeng~Sun,
        and~Yingfei~Xiong%
\IEEEcompsocitemizethanks{%
\IEEEcompsocthanksitem Zhechong Huang, Zhao Zhang, and Yingfei Xiong are with the School of Computer Science, Peking University, Beijing, China. E-mail: \{willhuang, zhangzhao2019\}@stu.pku.edu.cn, xiongyf@pku.edu.cn.
\IEEEcompsocthanksitem Zeyu Sun is with the Institute of Software, Chinese Academy of Sciences, Beijing, China. E-mail: zeyu.zys@gmail.com.
\IEEEcompsocthanksitem Huifeng Sun is with Tencent, China. E-mail: shelon\_2008@126.com.}}

\markboth{IEEE Transactions on Software Engineering,~Vol.~XX, No.~XX, 2026}%
{Huang \MakeLowercase{\textit{et al.}}: SpecRL with Test-Based Completeness Rewards}

\IEEEtitleabstractindextext{%
\begin{abstract}
Specification synthesis asks a model to generate specifications and auxiliary annotations for an existing program. In modern software verification projects, specification accuracy is critical. 
A specification is the abstraction of a method's behavior, and callers are verified against the callee's specification rather than its implementation. As a result, the specification for a callee can strongly affect what properties callers can prove. 
However, verifier feedback alone is a poor training signal for specification synthesis: verification can prove that a specification is sound for the implementation, yet it cannot tell whether the specification is too weak.

We present \textbf{SpecRL}, a reinforcement learning framework that adds an empirical completeness signal to specification synthesis in Dafny. SpecRL constructs negative tests, or \emph{spectests}, from implementation-impossible input-output pairs that weak specifications such as \texttt{ensures true} may still admit.
During training, SpecRL rewards verified candidates according to the fraction of spectests their specifications reject, thereby ranking these candidates by how many implementation-impossible behaviors they rule out. 
On the out-of-distribution DafnyComp-Spec benchmark, the 7B SpecRL model improves verification success and completeness over supervised fine-tuning by 49.96\% and 26.46\%, respectively. 
These relative gains show that fine-grained spectest feedback improves both verifiability and specification accuracy.

\end{abstract}

}

\maketitle
\IEEEdisplaynontitleabstractindextext
\IEEEpeerreviewmaketitle

\section{Introduction}
\emergencystretch=1em

Formal verification proves that a program satisfies desired properties by checking it against a \emph{formal specification}~\cite{hoare1969axiomatic}.
In auto-active verification languages such as Dafny~\cite{leino2010dafny}, Frama-C/ACSL~\cite{baudin2021acsl}, and Verus~\cite{lattuada2023verus}, developers write preconditions and postconditions to describe method behavior, and auxiliary annotations such as loop invariants, assertions, and lemmas to help the verifier discharge proof obligations.
Writing these annotations by hand is tedious and error-prone~\cite{wen2024autospec}, motivating growing interest in LLM support for formal verification.
\
Recent work has targeted Dafny and Verus verification tasks, including auxiliary annotation generation~\cite{mugnier2024laurel, banerjee2026dafnypro, yang2025autoverus, aggarwal2025alphaverus}.
Related efforts also study loop invariant synthesis~\cite{chakraborty2023ranking, wu2024lam4inv, cao2025clause2inv} and other satisfiability modulo theories (SMT)-assisted proof-oriented programming~\cite{chakraborty2025neural}.

In this paper, we focus on \emph{specification synthesis}, a central task in formal verification.
Given an existing implementation with missing annotations, the goal is to generate specifications and auxiliary annotations so that the program verifies and the generated specifications accurately characterize the method's behavior.
Specifications serve a broader role than helping a method verify.
A specification is a machine-checked abstraction of a method's behavior~\cite{meyer1992applying, hatcliff2012behavioral}: developers read it to understand code, and verifiers use it to reason about method calls.
This matters in modern software verification projects because callers are verified against the callee's specification rather than its implementation body, making specification accuracy critical.
For example, a caller may rely on a callee to return the maximum element of an array. If the callee specification only says that the return value is no smaller than every array element, the caller cannot prove properties that require the returned value to be an array element, even when the callee implementation is correct.
Specification synthesis therefore needs to optimize for both local verifiability and accuracy: the ability to rule out behaviors the implementation cannot produce, such as returning \texttt{100} for input \texttt{[3,1,4,1,5]} in the example above.

Recent systems have made progress on automated formal verification, but how to train models to generate specifications that are both verifiable and accurate remains open.
Inference-time systems like AutoSpec~\cite{wen2024autospec}, Clover~\cite{sun2024clover}, and Assured~\cite{mirchev2024assured} improve outputs with prompting, static analysis, filtering, or repair. However, the model itself does not internalize much understanding of what makes a specification accurate.
Supervised fine-tuning approaches like SAFE~\cite{chen2025safe} and VeRuSyn~\cite{di2026verusyn} fine-tune on verified outputs, but supervised learning inherits the accuracy and coverage limits of its training data.
Reinforcement learning has improved code generation by using compiler or execution feedback~\cite{le2022coderl, dou2024stepcoder, gehring2025rlef}. ReForm~\cite{yan2025reform} further shows that verifier feedback can be used in training Dafny models.
However, verification feedback is coarse for specification synthesis. It tells whether the implementation satisfies a specification, but it cannot distinguish an accurate specification from a weak one such as \texttt{ensures true}: both can pass the verifier for the implementation.
\
Thus, a reward for specification synthesis should go beyond verification success and provide a finer signal for distinguishing accurate specifications from weak ones.

We propose \textbf{SpecRL}, a reinforcement learning framework for specification synthesis in Dafny.
SpecRL constructs \textbf{spectests}: specification-level negative tests derived from concrete program executions.
For each target method, SpecRL asks an LLM for concrete inputs, executes the implementation to observe the corresponding outputs, and mutates only those outputs. The resulting input-output pairs are impossible for the deterministic implementation: an accurate specification should reject them, while a weak specification may still admit them.
The fraction of spectests rejected by a candidate specification gives an empirical completeness signal for RL: among candidates that already pass verification, the reward favors specifications that rule out more impossible behaviors.
\
SpecRL uses spectests to define the reward and evaluate specification accuracy, while keeping inference simple: the trained model receives the stripped Dafny program and directly generates specifications and auxiliary annotations.

This paper makes three contributions:

\begin{enumerate}
    \item \textbf{An execution-backed spectest construction pipeline} that builds reusable spectest suites without requiring pre-existing tests. The pipeline turns observed executions into implementation-impossible input-output pairs that accurate specifications should reject.

    \item \textbf{A progressive RL reward for specification synthesis} that first requires a candidate specification to be compilable and verifiable with respect to the implementation, and then uses spectest rejection rate to distinguish accurate specifications from weak ones.

    \item \textbf{A comprehensive evaluation on three Dafny benchmarks enhanced with our spectest construction pipeline}. On these benchmarks, SpecRL improves both verifiability and specification accuracy over SFT and a ReForm-style RL baseline across four model backbones. Trained small models also remain competitive with much larger general-purpose LLMs on the out-of-distribution DafnyComp-Spec benchmark.
\end{enumerate}

\section{Background}
\label{sec:background}

\subsection{Verification Annotations in Dafny}
\label{sec:background:dafny}

Dafny~\cite{leino2010dafny} is a verification-aware language whose built-in verifier translates programs and annotations into verification conditions and discharges them with an SMT solver~\cite{demoura2008z3}.
These annotations fall into two categories:

\begin{itemize}
    \item \textbf{Specifications} (\textcolor{specblue}{\textbf{blue}}): \texttt{requires} clauses state preconditions on method inputs; \texttt{ensures} clauses state postconditions on both method inputs and outputs. Together they define the method-behavior abstraction.
    \item \textbf{Auxiliary annotations} (\textcolor{proofpurple}{\textbf{purple}}): loop invariants and assertions that help the verifier discharge proof obligations.
\end{itemize}

\noindent
Figure~\ref{fig:overview-example} shows a Dafny method \texttt{Max} that scans an array of natural numbers and returns the maximum element (or \texttt{-1} for an empty array).
The shown \textcolor{specblue}{\texttt{requires}} clause restricts the specification to non-empty arrays, and the \textcolor{specblue}{\texttt{ensures}} clause states that the returned value is at least every element.
The loop \textcolor{proofpurple}{\texttt{invariant}} is an auxiliary annotation that states a property maintained at every iteration, enabling the verifier to reason about the loop without unrolling it.
Taken together, these annotations specify that, for a non-empty array, \texttt{Max} returns a value no smaller than any array element.

\begin{figure}[t]
\centering
\resizebox{\columnwidth}{!}{%
\begin{tikzpicture}[
  x=1cm,
  y=0.95cm,
  code/.style={anchor=west, font=\footnotesize\ttfamily, inner sep=0pt},
  tag/.style={font=\tiny\sffamily\bfseries, fill=codebg, inner xsep=2pt, inner ysep=1pt},
]
  \filldraw[fill=codebg, draw=codegray!40, line width=0.4pt]
    (-0.12,0.24) rectangle (8.55,-6.16);

  \node[code] (l01) at (0,0) {\textbf{method} Max(a: array\ensuremath{<}nat\ensuremath{>}) \textbf{returns} (m: int)};
  \node[code] (l02) at (0,-0.36) {\ \ \textcolor{specblue}{\textbf{requires}} a.Length \ensuremath{>} 0};
  \node[code] (l03) at (0,-0.72) {\ \ \textcolor{specblue}{\textbf{ensures}} \textbf{forall} k ::};
  \node[code] (l04) at (0,-1.08) {\ \ \ \ 0 \ensuremath{\leq} k \ensuremath{<} a.Length \ensuremath{\Longrightarrow} m \ensuremath{\geq} a[k]};
  \node[code] (l05) at (0,-1.44) {\char`\{};
  \node[code] (l06) at (0,-1.80) {\ \ \textbf{if} a.Length == 0 \char`\{ \textbf{return} -1; \char`\}};
  \node[code] (l07) at (0,-2.16) {\ \ \textbf{var} i := 0;};
  \node[code] (l08) at (0,-2.52) {\ \ m := a[0];};
  \node[code] (l09) at (0,-2.88) {\ \ \textbf{while} i \ensuremath{<} a.Length};
  \node[code] (l10) at (0,-3.24) {\ \ \ \ \textcolor{proofpurple}{\textbf{invariant}} 0 \ensuremath{\leq} i \ensuremath{\leq} a.Length};
  \node[code] (l11) at (0,-3.60) {\ \ \ \ \textcolor{proofpurple}{\textbf{invariant}} \textbf{forall} k ::};
  \node[code] (l12) at (0,-3.96) {\ \ \ \ \ \ 0 \ensuremath{\leq} k \ensuremath{<} i \ensuremath{\Longrightarrow} m \ensuremath{\geq} a[k]};
  \node[code] (l13) at (0,-4.32) {\ \ \char`\{};
  \node[code] (l14) at (0,-4.68) {\ \ \ \ \textbf{if} a[i] \ensuremath{\geq} m \char`\{ m := a[i]; \char`\}};
  \node[code] (l15) at (0,-5.04) {\ \ \ \ i := i + 1;};
  \node[code] (l16) at (0,-5.40) {\ \ \char`\}};
  \node[code] (l17) at (0,-5.76) {\char`\}};

  \draw[specblue, rounded corners=1pt, line width=0.65pt]
    (-0.05,-0.16) rectangle (7.72,-1.30);
  \node[tag, text=specblue, anchor=east] at (7.68,-0.38) {Specifications};

  \draw[proofpurple, rounded corners=1pt, line width=0.65pt]
    (-0.05,-3.03) rectangle (7.72,-4.16);
  \node[tag, text=proofpurple, anchor=east] at (7.68,-3.26) {Auxiliary annotations};
\end{tikzpicture}%
}
\caption{A Dafny method \texttt{Max} with \textcolor{specblue}{specifications} and \textcolor{proofpurple}{auxiliary annotations}.}
\label{fig:overview-example}
\end{figure}

\subsection{Soundness and Completeness of Specifications}
\label{sec:background:cc}

A specification defines a set of allowed input-output behaviors: the precondition constrains the input, and the postcondition constrains the output for each valid input.
For a deterministic method, we write each behavior as an input-output pair $(x,y)$.
A Hoare triple $\{\mathit{Pre}\}\ P\ \{\mathit{Post}\}$ holds when every execution of $P$ that starts from an input satisfying $\mathit{Pre}(x)$ terminates with an output satisfying $\mathit{Post}(x,y)$.
As a behavior relation, the specification admits a behavior $(x,y)$ when the predicate $\mathit{Pre}(x) \Rightarrow \mathit{Post}(x,y)$ evaluates to true, and rejects it otherwise.
Any behavior is admitted when the precondition is false or the postcondition is true.

Following prior work on specification evaluation~\cite{lahiri2024evaluating}, we view a specification through the set of behaviors it admits.
\
We write $\sem{S}$ for the behaviors permitted by a specification $S$, and $\sem{P}$ for the behaviors exhibited by the program $P$.
In this set-based view, soundness and completeness correspond to two directions of approximation: a sound specification must over-approximate the program's behaviors, while a complete specification should avoid admitting behaviors outside the program.
Soundness and completeness are program-relative notions: they measure how tightly a specification describes the implementation, rather than inferring intent that is not present in the code.
The relation between $\sem{S}$ and $\sem{P}$ can be characterized by these two properties:

\begin{itemize}
    \item \textbf{Soundness.} $\sem{P} \subseteq \sem{S}$: every behavior that the program can produce satisfies the specification.
    \item \textbf{Completeness.} $\sem{S} \subseteq \sem{P}$: every behavior admitted by the specification can be produced by the program.
\end{itemize}

\noindent
When $\sem{S} = \sem{P}$, the specification is both sound and complete.
However, Dafny verification \emph{only establishes the soundness direction}: it checks that the implementation satisfies the specification, not that the specification excludes all implementation-impossible behaviors.
As a result, specifications in verified datasets can be sound but incomplete.
\
A vacuous postcondition such as \texttt{ensures true} illustrates the problem: it is satisfied by any method implementation and can always pass verification while conveying no useful behavioral information.
The shown specification in Figure~\ref{fig:overview-example} is less vacuous but still incomplete in two ways: the postcondition is \emph{too weak} because it permits returning any sufficiently large number, not just the maximum, and the precondition is \emph{too strong} because \texttt{requires a.Length > 0} excludes empty arrays, for which the method has well-defined behavior.
By contrast, a \emph{complete} specification for \texttt{Max} would characterize both cases handled by the implementation, for example:
\begin{lstlisting}
ensures a.Length == 0 ==> m == -1
ensures a.Length > 0 ==> m in a[..]
ensures forall k :: 0 <= k < a.Length ==> m >= a[k]
\end{lstlisting}
\noindent
These clauses say that the method returns \texttt{-1} on empty arrays and otherwise an array element that is no smaller than every other element. As a result, they rule out spurious behaviors such as returning \texttt{100} for \texttt{[3,1,4,1,5]} while still admitting every behavior that the program can actually produce.

\subsection{Reinforcement Learning for Model Training}
\label{sec:background:rl}

Supervised fine-tuning (SFT) trains a language model to imitate reference outputs and is thus inherently limited by the accuracy and coverage of its training data~\cite{ouyang2022training}.
Reinforcement learning (RL) overcomes this limitation by letting the model explore: it samples diverse outputs and receives scores from a \emph{reward function} $r(x, y)$, reinforcing outputs that score well regardless of whether they appear in the training set.
RL also allows the model to exploit \emph{non-differentiable} feedback signals such as compiler or verifier verdicts, which SFT cannot directly incorporate.

We use \textbf{Group Relative Policy Optimization} (GRPO)~\cite{shao2024deepseekmath} for training.
In each iteration, for a given prompt $x$, GRPO samples a group of $G$ responses and scores them with $r(x, y)$.
Each response's \emph{advantage} is computed relative to the group:
\begin{equation}
\label{eq:grpo-advantage}
\hat{A}_i = \frac{r(x, y_i) - \mathrm{mean}(\{r(x, y_j)\}_{j=1}^{G})}{\mathrm{std}(\{r(x, y_j)\}_{j=1}^{G})}
\end{equation}
The model's parameters are then updated to increase the likelihood of responses with positive advantage and decrease the likelihood of those with negative advantage. Over successive iterations, the policy gradually shifts toward generating outputs that receive higher rewards.

\section{Method}
\label{sec:method}

This section defines the specification-synthesis task, gives an overview of SpecRL, and then describes its offline spectest construction and online RL reward design.

\subsection{Problem Setting}
\label{sec:overview:problem}

The input to SpecRL is a fully implemented Dafny method after its specifications and auxiliary annotations have been removed, which is called a \emph{stripped program}. We focus on deterministic methods, so running a method on the same input produces the same observed output. The model must add preconditions, postconditions, and auxiliary annotations such as loop invariants, while preserving the original method signature and implementation body. Dafny then checks whether the resulting annotated program verifies.

\subsection{Overview of SpecRL}
\label{sec:overview}

\begin{figure*}[t]
\centering
\includegraphics[width=0.98\textwidth]{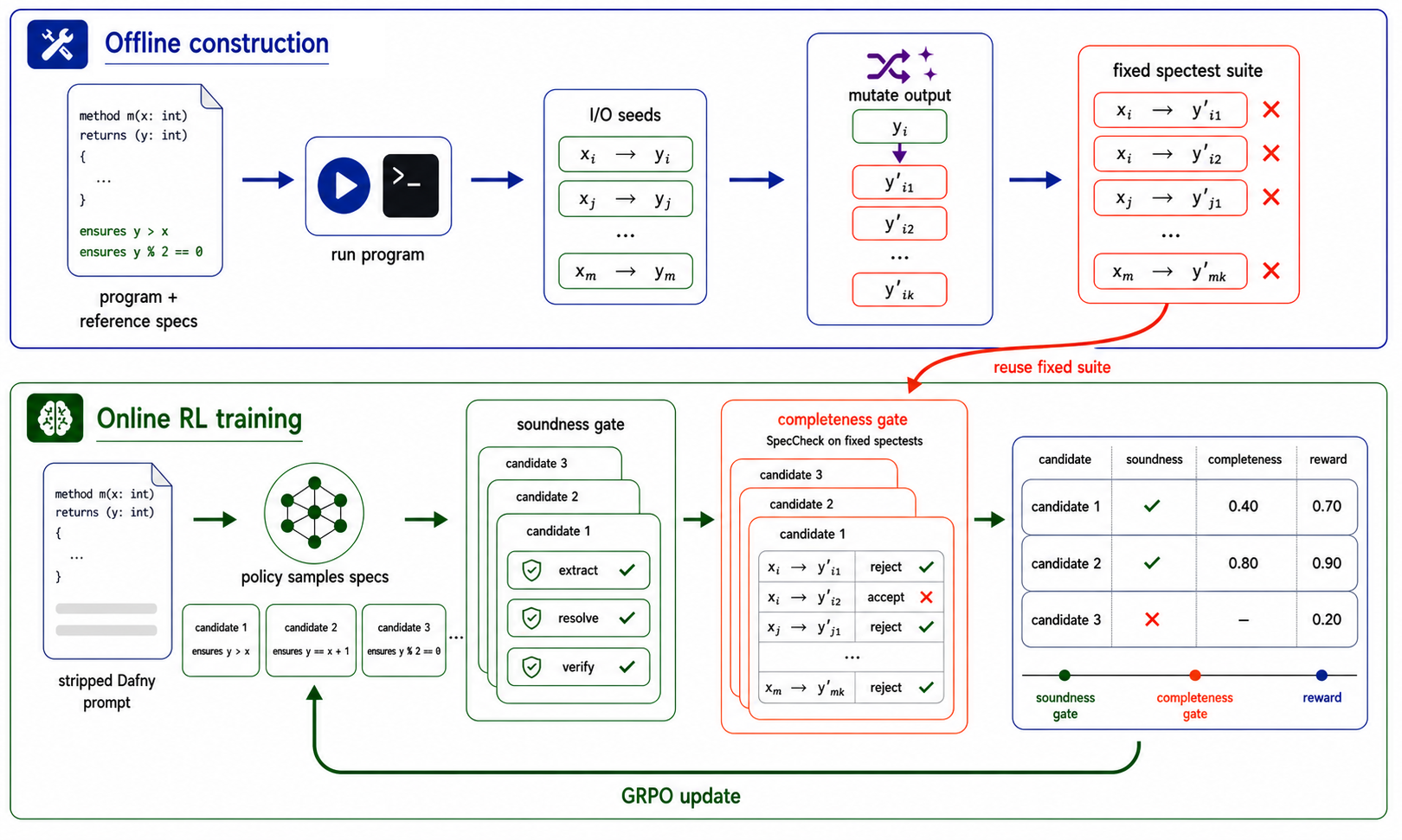}
\caption{High-level SpecRL workflow. Offline construction builds fixed spectest suites from execution-backed inputs and output mutations. Online RL reuses these suites after the extraction, compilation, and verification gates to score candidate annotations.}
\label{fig:specrl-overall}
\end{figure*}

Figure~\ref{fig:specrl-overall} shows the overall SpecRL architecture, which consists of offline spectest construction and online RL training. We use the running example below to illustrate how the fixed spectests connect these two parts.

\parhead{Offline}
Consider the \texttt{Max} method from Figure~\ref{fig:overview-example}. SpecRL uses the annotated program in the original dataset, executes the method on a concrete input to obtain the return value, and then constructs a \emph{spectest} by pairing the input with a mutated wrong-output value. A few spectests for \texttt{Max} are:

\begin{center}
\footnotesize
\setlength{\tabcolsep}{2.5pt}
\begin{tabular*}{\columnwidth}{@{\extracolsep{\fill}}llll@{}}
\toprule
\textbf{Input} & \textbf{Observed} & \textbf{Mutated} & \textbf{Case} \\
\midrule
\texttt{[3,1,4,1,5]} & \texttt{5} & \texttt{14}  & sum \\
\texttt{[3,1,4,1,5]} & \texttt{5} & \texttt{100} & large \\
\texttt{[3,1,4,1,5]} & \texttt{5} & \texttt{3}   & non-max \\
\texttt{[]}          & \texttt{-1} & \texttt{99} & empty \\
\bottomrule
\end{tabular*}
\end{center}

SpecRL chooses these output mutations to target plausible specification weaknesses, such as missing empty-array behavior or omitting the condition that the returned value should be an actual element. Section~\ref{sec:design-rationale} gives the input-output (I/O) space view behind this construction.

\parhead{Online}
During online RL, the model receives the stripped program and samples candidate annotations. Suppose it generates the following two specifications, both of which can pass Dafny verification given proper auxiliary annotations:

\noindent
\texttt{Spec\_weak}:
\begin{lstlisting}
requires a.Length > 0
ensures forall k :: 0 <= k < a.Length ==> m >= a[k]
\end{lstlisting}

\noindent\texttt{Spec\_strong}:
\begin{lstlisting}
ensures a.Length > 0 ==> m in a[..]
ensures forall k :: 0 <= k < a.Length ==> m >= a[k]
\end{lstlisting}

\noindent
We say that a specification \emph{rejects} a spectest when it evaluates to \texttt{false} on that input-output pair.
SpecRL computes the empirical completeness score as the fraction of spectests rejected by the candidate specification. Against the four spectests listed above, \texttt{Spec\_weak} rejects only one, so its score is \texttt{1/4 = 0.25}. \texttt{Spec\_strong} rejects three and receives \texttt{3/4 = 0.75}. The two candidates are both verifiable, but the spectest reward gives RL a gradient toward the more accurate one. Across RL updates, this signal encourages the policy to add missing clauses, such as the empty-array behavior, until the generated specification can reject all spectests.

\subsection{Offline Spectest Construction}
\label{sec:speccheck}
Given a Dafny file containing a program together with its specification, SpecRL produces a self-contained test file for Dafny's testing framework\footnote{\texttt{dafny test}: runs all methods annotated with the \texttt{\symbol{123}:test\symbol{125}}.}.
The construction combines symbolic transformations (Steps~1--3 and 6) with LLM-based generation (Steps~4--5 and 7), using the LLM only where candidate inputs, wrong-output mutations, or weakness analyses are needed.

\subsubsection{Step 1: Method Extraction}
\label{sec:speccheck:step1}

A Dafny source file may contain several methods as well as helper functions, predicates, classes, and datatypes.
Step~1 selects one primary executable method as the spectest target; the other auxiliary declarations remain part of the program context but are not themselves used as spectest targets.
This choice keeps the later input generation, output mutation, and reward computation tied to one method contract instead of mixing the contracts of several methods.

\subsubsection{Step 2: \textnormal{\texttt{SpecCheck}} Predicate Construction}
\label{sec:speccheck:step2}

From the selected method, Step~2 constructs a \texttt{predicate SpecCheck} that evaluates its specification.
\texttt{SpecCheck}'s parameters are the method inputs and return values.
Each \texttt{ensures} clause is conjoined with \texttt{\&\&} to form the predicate body.
If the method has \texttt{requires} clauses, they are folded into the body using implication (\texttt{==>}), so that the predicate evaluates to \texttt{true} whenever the precondition is not met.
For example, given:

\begin{lstlisting}
method M(x: int) returns (y: int)
  requires R1
  requires R2
  ensures E1
  ensures E2
\end{lstlisting}

\noindent
Step~2 produces:

\begin{lstlisting}
predicate SpecCheck(x: int, y: int)
{ (R1 && R2) ==> (E1 && E2) }
\end{lstlisting}

\subsubsection{Step 3: Predicate Compilation Check}
\label{sec:speccheck:step3}

Step~3 compiles the source file together with \texttt{SpecCheck} to determine whether it can be executed at runtime.
Methods whose specification clauses do not yield compilable predicates, such as clauses referencing ghost functions or unbounded quantifiers, are excluded from the test suite.

\subsubsection{Step 4: Input-Output Pair Generation}
\label{sec:speccheck:step4}

Step~4 derives input-output pairs for the target method.
An LLM proposes diverse candidate inputs that cover different execution paths, boundary cases, and cases near or outside the boundary of \texttt{requires} clauses.
The LLM supplies only inputs; SpecRL obtains the outputs by executing those inputs on the target method.
If execution fails, Step~4 enters a \emph{reflect-and-fix} loop for up to three rounds, feeding the failure back to the LLM so it can revise the input.

\subsubsection{Step 5: Negative Mutation Generation}
\label{sec:speccheck:step5}

Step~5 asks the LLM to propose wrong-output values for each input-output pair.
These values target boundary, off-by-one, corner-case, and other plausible outputs that a weak specification may mistakenly admit.
If a generated test fails to compile or execute, Step~5 uses the same reflect-and-fix loop as Step~4.
Each spectest follows a structure like:
\begin{lstlisting}
method {:test} Neg_IO1_1() {
  var a := new nat[] [3, 1, 4, 1, 5];
  var m := Max(a);
  m := 14;             // mutated output (sum)
  expect !SpecCheck(a, m);
}
\end{lstlisting}

\noindent

The target methods are deterministic, so the mutated pairs are impossible relative to the implementation.
Yet they are chosen to target plausible specification weaknesses, so a weak specification may still fail to reject them.

\subsubsection{Step 6: Assembly and Runtime Evaluation}
\label{sec:speccheck:step6}

Step~6 assembles a Dafny file containing the original source file, the \texttt{SpecCheck} predicate from Step~2, and all spectests from Step~5.
Supplementary Appendix~C gives the concrete file shape for the \texttt{Max} example.
Executing this file gives the \emph{spectest rejection rate}: the fraction of spectests rejected by the specification, which serves as its empirical completeness score.

\subsubsection{Step 7: Iterative Suite Enhancement}
\label{sec:iterate-enhance}
A high spectest rejection rate, especially when all spectests are rejected, is ambiguous: the current specification may already be accurate, or the suite may simply have missed its weaknesses.
To check for the second case, Step~7 performs an enhancement after Steps~1--6.
At each enhancement iteration, it reuses the current test results, specification, and spectest suite, and asks an LLM to analyze which implementation-impossible behaviors may still escape.
The LLM may either conclude that the suite is adequate or output concrete weakness descriptions.
If weaknesses are reported, SpecRL inserts them into the prompts of Steps~4 and~5 and reruns Steps~4--6 to build an enhanced suite.
If the LLM judges the suite adequate, Step~7 stops without further enhancement.

\subsubsection{Design Rationale: I/O Space Perspective}
\label{sec:design-rationale}

\begin{figure}[t]
  \centering
  \includegraphics[width=\columnwidth]{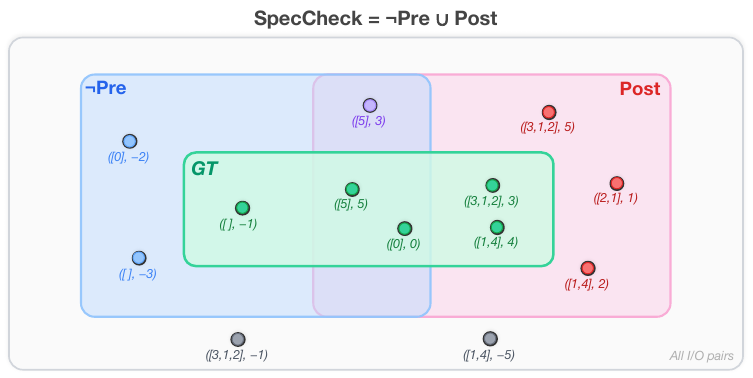}
  \caption{A Venn-style diagram in I/O space for method \texttt{Max} under another specification \texttt{requires a.Length > 1; ensures m >= 0}. The blue region is $\neg\mathrm{Pre}$, the pink region is $\mathrm{Post}$, and their union is the $\mathrm{SpecCheck}$. The green region is $\mathrm{GT}$, the ground-truth set of implementation behaviors.}
  \label{fig:spectest-venn}
\end{figure}

We use the same set-theoretic view as Section~\ref{sec:background:cc}: programs and specifications are treated as sets of input-output behaviors.
Figure~\ref{fig:spectest-venn} instantiates this view, where each point is an input-output pair $(x, y)$.
Let $\mathit{Pre}$ and $\mathit{Post}$ denote the conjunction of all \texttt{requires} and \texttt{ensures} clauses.
We write $\mathrm{Pre}$ and $\mathrm{Post}$ for the sets of behaviors satisfying these two propositions, $\neg\mathrm{Pre}$ for the complement of $\mathrm{Pre}$, and $\mathrm{GT}$ for the set of behaviors exhibited by the implementation.

Since the specification equals the predicate $\mathit{Pre} \Rightarrow \mathit{Post}$, it accepts behaviors in $\neg\mathrm{Pre} \cup \mathrm{Post}$. We write $\mathrm{SpecCheck}$ for this region.
Verification establishes soundness, which means $\mathrm{GT} \subseteq \mathrm{SpecCheck}$.
The gap $\mathrm{SpecCheck} \setminus \mathrm{GT}$ contains the implementation-impossible behaviors that this weak specification still permits.
For example, both behaviors $([5], 3)$ and $([3,1,2], 5)$ in Figure~\ref{fig:spectest-venn} lie in this gap.

For a fixed test suite, a higher spectest rejection rate means that fewer behaviors remain in the gap, so the specification appears more complete. 
During suite construction, however, the preference is reversed. Behaviors outside $\mathrm{SpecCheck}$ are trivially rejected and provide no further optimization signal. So a lower spectest rejection rate means that more mutated behaviors still land in the gap, making the suite more effective at exposing weaknesses. 

This is our design rationale for Steps 4, 5 and 7.
Step~4 starts from diverse behaviors in $\mathrm{GT}$ while trying to propose inputs outside the precondition.
Step~5 mutates the output to produce behaviors that the postcondition still permits, so the mutated pair leaves $\mathrm{GT}$ by construction.
The spectests are outside $\mathrm{GT}$ but likely to stay in $\mathrm{SpecCheck}=\neg\mathrm{Pre} \cup \mathrm{Post}$ through one of two routes: a mutated output that's still within $\mathrm{Post}$, or an input already outside $\mathrm{Pre}$. Step 7 further increases the proportion of behaviors within this gap by proposing more weakness-targeted behaviors.

\subsection{Online RL Optimization with Spectest Reward}
\label{sec:spectest-reward}

With the offline spectest suites fixed, the online phase trains a policy over stripped Dafny programs.
At each training step, SpecRL selects one training program and samples a group of candidate annotations.
Each candidate then passes through the progressive reward for optimization: SpecRL first extracts Dafny code from the model output, checks compilation, verifies the annotated program, and then runs the spectest suite to score the candidate specifications.
GRPO updates the policy from the group-relative rewards, gradually favoring specifications that make the program verify and reject more implementation-impossible behaviors.

\subsubsection{Reward Design}
Each sampled candidate is scored with a four-stage progressive reward. The first three stages check whether the sampled output is usable Dafny code: extraction, compilation, and verification. The fourth stage uses the offline spectest suite to distinguish weak specifications from more accurate ones among verified candidates. Formally, the reward $r$ is
\begin{equation}
\label{eq:reward}
r = r_{\mathrm{extract}} + r_{\mathrm{compile}} + r_{\mathrm{verify}} + r_{\mathrm{spectest}}
\end{equation}
where each component is defined as follows.

\parhead{Stage 1: Extraction ($r_{\mathrm{extract}}$)}
The LLM output is parsed to extract a valid Dafny file with generated annotations.
If extraction succeeds, $r_{\mathrm{extract}} = 0.05$; otherwise, $r_{\mathrm{extract}} = 0$ and all subsequent stages are skipped.

\parhead{Stage 2: Compilation ($r_{\mathrm{compile}}$)}
The extracted file is compiled to ensure that the annotated program is syntactically well formed and type-correct.
If compilation succeeds, $r_{\mathrm{compile}} = 0.15$; otherwise, $r_{\mathrm{compile}} = 0$ and later stages are skipped.

\parhead{Stage 3: Verification ($r_{\mathrm{verify}}$)}
The method is submitted to the Dafny verifier to check that the specification is provably consistent with the implementation.
If verification succeeds, $r_{\mathrm{verify}} = 0.30$; otherwise, $r_{\mathrm{verify}} = 0$ and Stage~4 is skipped.

\parhead{Stage 4: Spectest reward ($r_{\mathrm{spectest}}$)}
This stage supplies the empirical completeness signal from offline spectest construction (Section~\ref{sec:speccheck}).
We gate it on successful verification so that spectest rejection is rewarded only among candidates that pass Dafny verification.
Given a candidate that passes verification, we construct a new \texttt{SpecCheck} predicate from it and run the pre-built offline spectest suite:
\begin{equation}
\label{eq:spectest-reward}
r_{\mathrm{spectest}} = 0.50 \times \frac{|\{t \in \mathcal{T}_{\mathrm{neg}} \mid t \text{ is rejected}\}|}{|\mathcal{T}_{\mathrm{neg}}|}
\end{equation}
where $\mathcal{T}_{\mathrm{neg}}$ is the spectest suite.
This stage asks only how many challenging mutated behaviors the candidate specification rejects.
A candidate that rejects all spectests receives the full $0.50$ reward, while a weak specification that rejects few receives a score close to $0$. 
For example, rejecting 6 out of 10 spectests receives $0.50 \times 6/10 = 0.30$ from this stage.

Following ReForm's progressive reward design~\cite{yan2025reform}, we keep the stage weights fixed: $0.05$ for extraction, $0.15$ for compilation, $0.30$ for verification, and $0.50$ for the accuracy.
The total reward ranges from $0$ (unparseable output) to $1.0$ (a verified specification that rejects all spectests). The first three stages provide coarse verification guidance early in training. Stage~4 then separates weak specifications from accurate ones among verifiable candidates.

\subsubsection{Reward shaping safeguards}
To prevent reward hacking, we apply two safeguards.
First, we enforce \emph{task preservation}: candidates receive $r = 0$ if they contain \texttt{assume} statements or modify the original implementation, because such changes bypass the intended verification task.

Second, we penalize \emph{trivial specifications}: candidates that emit vacuous annotations such as \texttt{ensures true} or \texttt{requires false} also receive $r = 0$.
This addresses a local optimum: a vacuous but verifiable specification can score higher than a meaningful specification attempt that fails verification.
Without the penalty, the policy may learn to prefer such vacuous annotations and require many updates to rediscover specifications that are both informative and verifiable.
Penalizing these cases pushes exploration toward nontrivial specifications.

\section{Evaluation}

We conduct a comprehensive evaluation of SpecRL by investigating the following research questions:
\begin{itemize}
  \item \textbf{RQ1}: Does SpecRL improve specification accuracy and verification success rate over baseline methods?
  \item \textbf{RQ2}: How does each reward component contribute to the overall performance?
  \item \textbf{RQ3}: Can task-specific RL make small models competitive with much larger general-purpose LLMs?
  \item \textbf{RQ4}: What cost does SpecRL add for RL training?
\end{itemize}

\subsection{Experimental Setup}
\label{sec:eval-setup}

\subsubsection{Benchmarks}
We construct three specification-generation benchmarks from existing Dafny datasets: \textbf{Py2Dfy-Spec} from Py2Dfy~\cite{yan2025reform}, \textbf{DafnyComp-Spec} from DafnyComp~\cite{yan2025reform}, and \textbf{DafnyBench-Spec} from DafnyBench~\cite{misu2024dafnybench}.
\begin{itemize}
  \item \textbf{Py2Dfy-Spec}: Py2Dfy contains Python programs translated into Dafny, including both single-method and multi-method programs. Its retained set is large enough for RL, so we randomly select 5\% of programs as the validation set, with the rest as the training set.
  \item \textbf{DafnyComp-Spec}: DafnyComp is derived from contest problems translated into Dafny. Its programs are longer and often require cross-method specification reasoning, making it closer to a realistic verification workload. We use it as the main evaluation set.
  \item \textbf{DafnyBench-Spec}: DafnyBench is a classic Dafny benchmark, but most of its programs are short, simple, and single-method. In our preliminary study, existing models can already produce accurate specifications for many cases. To reduce evaluation cost and avoid easy-case inflation, we build the additional evaluation set from the longest quartile of the retained programs.
\end{itemize}

For each source dataset, we apply the spectest construction pipeline from Section~\ref{sec:speccheck} offline and keep only programs whose corresponding spectest suite can be constructed.
We use DeepSeek-V3~\cite{deepseek2025v3} for all LLM-based stages except iterative test enhancement, for which we use GPT-5.1~\cite{openai2026gpt51}.
Table~\ref{tab:dataset-overview} summarizes the resulting benchmark sizes, and Supplementary Appendix~B provides detailed statistics for the source and filtered datasets.

\begin{table}[t]
\centering
\caption{Dataset overview after offline spectest construction. Avg. denotes the average number of spectests per program.}
\label{tab:dataset-overview}
\small
\setlength{\tabcolsep}{3pt}
\begin{tabular*}{\columnwidth}{@{\extracolsep{\fill}}lrrr@{}}
\toprule
\textbf{Benchmark} & \textbf{Programs} & \textbf{Spectests} & \textbf{Avg.} \\
\midrule
Py2Dfy-Spec & 4,663 & 54,284 & 11.64 \\
DafnyComp-Spec & 232 & 2,913 & 12.56 \\
DafnyBench-Spec & 107 & 1,059 & 9.90 \\
\bottomrule
\end{tabular*}
\end{table}

\subsubsection{Metrics}
In the evaluation, we report four metrics:
\textbf{Compilable} (Dafny compiles successfully),
\textbf{Verifiable} (passes Dafny verification),
\textbf{Completeness} (empirical, the spectest rejection rate), and
\textbf{Spec-Superiority}.
Spec-Superiority~\cite{yan2025reform} is a reference-based metric and uses the existing reference specification provided by the dataset.
Specifically, we ask Dafny to prove that the generated specification implies the reference specification, written as $\mathit{spec}_{\mathrm{gen}} \Rightarrow \mathit{spec}_{\mathrm{ref}}$.
Since passing this implication check indicates that the generated specification is logically at least as strong as the dataset reference, Spec-Superiority provides a third-party signal of specification accuracy.

\subsubsection{Models and Baselines}
We use four Qwen2.5~\cite{qwen2025qwen25} backbones: 0.5B, 1.5B, 3B, and 7B. ReForm releases SFT checkpoints at these scales, giving us Dafny-adapted starting points and allowing a controlled comparison from the same initialization.
For each backbone, we compare the SFT checkpoint with two RL variants trained on the Py2Dfy-Spec:
\begin{itemize}
  \item \textbf{SFT}: the ReForm SFT checkpoint, without any RL.
  \item \textbf{ReForm-style RL}: a reference-based RL baseline. It uses the same extraction, compilation, and verification stages as SpecRL, but replaces Stage~4 with a binary reward based on the Spec-Superiority criterion.
  \item \textbf{SpecRL}: RL with our spectest reward, where Stage~4 scores candidates by the spectest rejection rate.
\end{itemize}
For brevity, we refer to ReForm-style RL as ReForm in the result tables and analysis below.

\subsubsection{Implementation Details}
For both RL variants, we use GRPO~\citep{shao2024deepseekmath} implemented in veRL~\citep{sheng2025verl}. We follow ReForm for the learning rate and sampling temperature and adopt standard veRL/GRPO settings for the remaining optimizer hyperparameters.
Concretely, we use a learning rate of $1\times10^{-5}$, temperature 1.0, group size $n{=}8$, Kullback--Leibler (KL) coefficient $0.001$, and Proximal Policy Optimization (PPO) clip ratio $0.2$, and train for 20 epochs. Batch size is adjusted only by model scale to fit GPU memory: for 0.5B and 1.5B, it is 512 (mini-batch 128); for 3B and 7B, it is 128 (mini-batch 32).

For both RL methods, we save one checkpoint every 10 training steps and select the checkpoint with the highest average reward on the Py2Dfy-Spec validation split. DafnyComp-Spec and DafnyBench-Spec are evaluated only after the checkpoint selection.

During inference, we sample 8 candidates per program without any pruning. Both \textbf{pass@1} and \textbf{pass@8} are computed from this 8-sample set. \textbf{pass@1} is estimated as the average single-sample success rate over these 8 candidates, and \textbf{pass@8} is the success rate over the full set.

\subsection{Results and Analysis}
\label{sec:eval-results}

The main results are reported on DafnyComp-Spec, which is not used for RL training or checkpoint selection. We compare SFT, ReForm, and SpecRL across four Qwen2.5 backbones.

\begin{table*}[t]
\centering
\caption{Main results on DafnyComp-Spec (\%). \textbf{Bold} marks the best value per model size. Parentheses report the relative change from the SFT checkpoint, in which \emph{n/a} means the SFT baseline is 0. }
\label{tab:main-results}
\small
\setlength{\tabcolsep}{2.2pt}
\renewcommand{\arraystretch}{1.08}
\resizebox{\textwidth}{!}{%
\begin{tabular}{@{}l@{\hspace{0.8em}}l@{\hspace{0.8em}}*{8}{c}@{}}
\toprule
\multirow{2}{*}{\raisebox{-0.65ex}{\textbf{\shortstack{Model\\size}}}} & \multirow{2}{*}{\raisebox{-0.35ex}{\textbf{Method}}} & \multicolumn{2}{c}{\textbf{Compilable}} & \multicolumn{2}{c}{\textbf{Verifiable}} & \multicolumn{2}{c}{\textbf{Completeness}} & \multicolumn{2}{c}{\textbf{Spec-Superiority}} \\
\cmidrule(lr){3-4} \cmidrule(lr){5-6} \cmidrule(lr){7-8} \cmidrule(lr){9-10}
& & pass@1 & pass@8 & pass@1 & pass@8 & pass@1 & pass@8 & pass@1 & pass@8 \\
\midrule
\multirow{3}{*}{0.5B} & SFT    & 60.08 & 70.69 & 1.19 & 3.02 & 16.73 & 19.68 & 0.00 & 0.00 \\
& ReForm & \textbf{77.21} (+28.5\%) & \textbf{81.90} (+15.9\%) & 2.86 (+140.3\%) & 5.17 (+71.2\%) & 16.96 (+1.4\%) & 19.27 (-2.1\%) & 0.16 (\emph{n/a}) & \textbf{1.29} (\emph{n/a}) \\
& SpecRL & 76.51 (+27.3\%) & 81.03 (+14.6\%) & \textbf{3.61} (+203.4\%) & \textbf{6.03} (+99.7\%) & \textbf{18.99} (+13.5\%) & \textbf{21.10} (+7.2\%) & \textbf{0.22} (\emph{n/a}) & \textbf{1.29} (\emph{n/a}) \\
\midrule
\multirow{3}{*}{1.5B} & SFT    & 66.33 & 78.02 & 1.72 & 4.31 & 17.80 & 20.94 & 0.11 & 0.43 \\
& ReForm & 77.96 (+17.5\%) & 82.33 (+5.5\%) & 3.07 (+78.5\%) & 5.60 (+29.9\%) & 19.51 (+9.6\%) & 22.17 (+5.9\%) & 0.43 (+290.9\%) & 2.59 (+502.3\%) \\
& SpecRL & \textbf{78.02} (+17.6\%) & \textbf{82.76} (+6.1\%) & \textbf{4.42} (+157.0\%) & \textbf{7.33} (+70.1\%) & \textbf{22.24} (+24.9\%) & \textbf{24.71} (+18.0\%) & \textbf{0.59} (+436.4\%) & \textbf{3.45} (+702.3\%) \\
\midrule
\multirow{3}{*}{3B} & SFT    & 69.61 & 81.90 & 3.77 & 9.05 & 17.14 & 20.17 & 2.32 & 3.45 \\
& ReForm & 78.45 (+12.7\%) & 83.19 (+1.6\%) & 6.63 (+75.9\%) & 12.07 (+33.4\%) & 17.82 (+4.0\%) & 20.25 (+0.4\%) & 2.59 (+11.6\%) & 4.74 (+37.4\%) \\
& SpecRL & \textbf{79.58} (+14.3\%) & \textbf{83.62} (+2.1\%) & \textbf{8.03} (+113.0\%) & \textbf{13.36} (+47.6\%) & \textbf{22.69} (+32.4\%) & \textbf{25.21} (+25.0\%) & \textbf{3.45} (+48.7\%) & \textbf{5.60} (+62.3\%) \\
\midrule
\multirow{3}{*}{7B} & SFT    & 67.03 & 78.88 & 5.44 & 11.21 & 18.15 & 21.35 & 2.93 & 6.47 \\
& ReForm & 78.02 (+16.4\%) & 82.33 (+4.4\%) & 7.81 (+43.6\%) & 14.22 (+26.9\%) & 17.58 (-3.1\%) & 19.98 (-6.4\%) & 3.02 (+3.1\%) & 7.76 (+19.9\%) \\
& SpecRL & \textbf{78.50} (+17.1\%) & \textbf{83.62} (+6.0\%) & \textbf{10.40} (+91.2\%) & \textbf{16.81} (+50.0\%) & \textbf{24.30} (+33.9\%) & \textbf{27.00} (+26.5\%) & \textbf{5.87} (+100.3\%) & \textbf{9.05} (+39.9\%) \\
\bottomrule
\end{tabular}%
}
\end{table*}

\subsubsection{Main Results}
RQ1 compares SFT, ReForm, and SpecRL across four Qwen2.5 backbones on DafnyComp-Spec under both pass@1 and pass@8 evaluation protocols (Table~\ref{tab:main-results}). We focus on the two main outcome metrics, Completeness and Verifiable, and use Compilable and Spec-Superiority as auxiliary evidence.

\parhead{Completeness}
SpecRL offers the largest Completeness gains at every model size, and the advantage widens as the backbone grows. Relative to SFT, SpecRL improves Completeness by +13.5\% at 0.5B and +33.9\% at 7B under pass@1; pass@8 follows the same pattern, rising from +7.2\% to +26.5\%. ReForm's Completeness gains are smaller and even turn negative at 7B ($-$3.1\% pass@1, $-$6.4\% pass@8).
This matches the reward design: a binary reference reward stops changing once the implication check succeeds, whereas spectest rejection continues to favor more accurate specifications.

\parhead{Verifiable}
Both RL methods guide the policy toward verifiability, but SpecRL still performs better: it improves Verifiable over SFT more consistently than ReForm. At 7B, SpecRL improves Verifiable by +91.2\% under pass@1 and +50.0\% under pass@8, compared with ReForm's +43.6\% and +26.9\%.
This is because verification is modular: callers rely on callee specifications, so a weak callee specification can make an otherwise valid caller unverifiable.
By pushing the policy toward more accurate specifications, SpecRL also improves full-program verification.

\parhead{Auxiliary evidence from Compilable and Spec-Superiority}
Compilable is similar between the two RL methods, and both improve over SFT.
On Spec-Superiority, SpecRL is higher than ReForm in every pass@1 result and is higher or tied in every pass@8 result; at 7B, SpecRL scores 5.87 versus ReForm's 3.02 under pass@1, and 9.05 versus 7.76 under pass@8.
Since Spec-Superiority also reflects specification accuracy, this reinforces the previous results: SpecRL improves specification accuracy while also improving Verifiable.

\parhead{DafnyBench-Spec}
To check whether this effect carries to other Dafny benchmarks, we evaluate the 7B models on DafnyBench-Spec (Table~\ref{tab:dafnybench-hard}).
Relative to SFT, SpecRL improves Verifiable by +17.7\% under pass@1 and +19.5\% under pass@8, and Completeness by +20.3\% and +18.2\%.
Compared with ReForm, SpecRL is also higher on all four scores: +6.6\%/+22.9\% on Verifiable and +12.7\%/+10.7\% on Completeness for pass@1/pass@8.
These results provide additional evidence that the findings are not specific to DafnyComp-Spec.

\begin{table}[t]
\centering
\caption{DafnyBench-Spec results for 7B models (\%). \textbf{Bold} marks the best value in each column. Parentheses report the relative change from SFT.}
\label{tab:dafnybench-hard}
\small
\renewcommand{\arraystretch}{1.12}
\setlength{\tabcolsep}{2.2pt}
\resizebox{\columnwidth}{!}{%
\begin{tabular}{@{}lcccc@{}}
\toprule
\multirow{2}{*}{\raisebox{-0.35ex}{\textbf{Method}}} & \multicolumn{2}{c}{\textbf{Verifiable}} & \multicolumn{2}{c}{\textbf{Completeness}} \\
\cmidrule(lr){2-3} \cmidrule(lr){4-5}
& pass@1 & pass@8 & pass@1 & pass@8 \\
\midrule
SFT    & 22.43 & 33.64 & 13.75 & 20.44 \\
ReForm & 24.77 (+10.4\%) & 32.71 (-2.8\%) & 14.67 (+6.7\%) & 21.84 (+6.8\%) \\
SpecRL & \textbf{26.40} (+17.7\%) & \textbf{40.19} (+19.5\%) & \textbf{16.54} (+20.3\%) & \textbf{24.17} (+18.2\%) \\
\bottomrule
\end{tabular}%
}
\end{table}

\begin{findingbox}[Finding 1 (RQ1)]
SpecRL consistently improves Completeness over both baselines, which is accompanied by higher Verifiable scores due to modular verification.
This holds on both DafnyComp-Spec and DafnyBench-Spec, together showing that SpecRL helps train more accurate specifications.
\end{findingbox}

\subsubsection{Ablation Study}
RQ2 studies how each reward component contributes to the overall performance. We fix the Qwen2.5-1.5B backbone and compare SpecRL with variants that remove the spectest reward or the verification gate, plus the SFT checkpoint as a no-RL reference (Table~\ref{tab:ablation}). All RL variants use the same training data and training budget.
We use 1.5B for this ablation because each row requires a separate RL training run. The 0.5B model is often too weak to show stable reward effects, while 3B and 7B make it much more expensive.

\begin{table}[t]
\centering
\caption{Ablation study on Qwen2.5-1.5B (\%). SpecRL is the full method, each \emph{w/o} row removes one reward component, and SFT is a no-RL reference. Parentheses report the relative change from \emph{SpecRL (full)}.}
\label{tab:ablation}
\small
\renewcommand{\arraystretch}{1.12}
\setlength{\tabcolsep}{2.2pt}
\resizebox{\columnwidth}{!}{%
\begin{tabular}{@{}lcccc@{}}
\toprule
\multirow{2}{*}{\raisebox{-0.35ex}{\textbf{Method}}} & \multicolumn{2}{c}{\textbf{Verifiable}} & \multicolumn{2}{c}{\textbf{Completeness}} \\
\cmidrule(lr){2-3} \cmidrule(lr){4-5}
& pass@1 & pass@8 & pass@1 & pass@8 \\
\midrule
SpecRL (full)   & \textbf{4.42} & \textbf{7.33} & 22.24 & 24.71 \\
\midrule
w/o Spectest    & 4.15 (-6.1\%) & 6.90 (-5.9\%) & 19.11 (-14.1\%) & 21.79 (-11.8\%) \\
w/o Verify      & 1.56 (-64.7\%) & 2.59 (-64.7\%) & \textbf{29.42} (+32.3\%) & \textbf{36.27} (+46.8\%) \\
\addlinespace[2pt]
SFT (no RL)     & 1.72 (-61.1\%) & 4.31 (-41.2\%) & 17.80 (-20.0\%) & 20.94 (-15.3\%) \\
\bottomrule
\end{tabular}%
}
\end{table}

We ablate two components:
\begin{itemize}
  \item \textbf{w/o Spectest}: Remove Stage~4 ($r_{\mathrm{spectest}}$), and the reward becomes $r = r_{\mathrm{extract}} + r_{\mathrm{compile}} + r_{\mathrm{verify}}$. This tests the necessity of the completeness reward.

  \item \textbf{w/o Verify}: Remove Stage~3 ($r_{\mathrm{verify}}$), and the reward becomes $r = r_{\mathrm{extract}} + r_{\mathrm{compile}} + r_{\mathrm{spectest}}$. This tests whether the verification gate is needed before rewarding empirical completeness.
\end{itemize}

Compared with \emph{SpecRL (full)}, \emph{w/o Spectest} lowers Completeness by 14.1\% under pass@1 and 11.8\% under pass@8, and its Verifiable score also falls below the full method.
Without the spectest reward, RL still learns from extraction, compilation, and verification, but it loses the signal that separates weak specifications from more accurate ones.

\emph{w/o Verify} exhibits the opposite failure mode. It achieves by far the highest Completeness, but its Verifiable score drops even below SFT. The model learns to reject many spectests without preserving provability. These outputs may score well under the finite spectest suite, but they are not useful for verification. This result shows why the verification gate is needed before the empirical completeness reward is applied.

\begin{findingbox}[Finding 2 (RQ2)]
The spectest reward and verification gate play complementary roles. Spectests push the model toward more accurate specifications, while the verification gate keeps the generated specifications provable.
\end{findingbox}

\subsubsection{Model Comparison}
RQ3 asks how competitive task-specific RL makes small models compared with much larger general-purpose LLMs. We compare SpecRL (Qwen2.5-3B/7B) with representative LLMs on DafnyComp-Spec, using the same prompt template and sampling budget.

\begin{table}[t]
\centering
\caption{Comparison with general-purpose LLMs on DafnyComp-Spec (\%).}
\label{tab:llm-comparison}
\small
\renewcommand{\arraystretch}{1.12}
\begin{tabular*}{\columnwidth}{@{\extracolsep{\fill}}lcccc@{}}
\toprule
& \multicolumn{2}{c}{\textbf{Verifiable}} & \multicolumn{2}{c}{\textbf{Completeness}} \\
\cmidrule(lr){2-3} \cmidrule(lr){4-5}
\textbf{Model} & pass@1 & pass@8 & pass@1 & pass@8 \\
\midrule
\multicolumn{5}{@{}l}{\emph{Task-specific RL}} \\
SpecRL (Qwen2.5-3B) & 8.03 & 13.36 & 22.69 & 25.21 \\
SpecRL (Qwen2.5-7B) & 10.40 & 16.81 & 24.30 & 27.00 \\
\midrule
\multicolumn{5}{@{}l}{\emph{Open-source}} \\
Qwen3-235B~\cite{yang2025qwen3}         & 4.03 & 6.90 & 11.12 & 20.97 \\
MiniMax-M2~\cite{minimax2025m2}          & 2.53 & 14.22 & 4.87 & 16.33 \\
DeepSeek-V3.2~\cite{deepseek2025v32}      & 7.00 & 15.09 & 13.53 & 23.53 \\
\midrule
\multicolumn{5}{@{}l}{\emph{Closed-source}} \\
Qwen2.5-Max       & 4.15 & 6.90 & 10.28 & 18.42 \\
Qwen3-Max           & 9.70 & 15.95 & 18.78 & 24.53 \\
GPT-5.1                                   & 11.10 & 22.84 & 19.60 & 26.33 \\
\bottomrule
\end{tabular*}
\end{table}

SpecRL-3B has higher Completeness than all open-source baselines in Table~\ref{tab:llm-comparison}, and SpecRL-7B is higher than those baselines on both Verifiable and Completeness. SpecRL-7B also outperforms flagship models from its own family (Qwen2.5-Max) and the next generation (Qwen3-235B, Qwen3-Max) across both metrics.
Against the strongest closed-source model, SpecRL-7B achieves higher Completeness, while GPT-5.1 remains better on Verifiable.
The result suggests that frontier models still retain an edge in verification capability. We interpret this comparison cautiously because all benchmarks used here are public, so both open-source and closed-source general-purpose models may have seen related code during their pretraining.

\begin{findingbox}[Finding 3 (RQ3)]
On DafnyComp-Spec, task-specific RL makes small models highly competitive with much larger general-purpose models. SpecRL-7B achieves the highest Completeness, although GPT-5.1 remains better on Verifiable.
\end{findingbox}
\subsubsection{Cost Analysis}
\label{sec:cost-analysis}
RQ4 analyzes the extra cost introduced by SpecRL's training framework. This cost has two parts: offline spectest construction and online GRPO training. 
The offline spectest suites are built once and reused for both RL and evaluation. Across the three benchmark suites used in this paper, spectest construction covers 5,002 programs and produces 58,256 spectests, with about \$250 in API cost. 

\begin{table}[t]
\centering
\caption{GRPO training cost for SpecRL runs with group size 8.}
\label{tab:training-cost}
\small
\renewcommand{\arraystretch}{1.12}
\begin{tabular*}{\columnwidth}{@{\extracolsep{\fill}}lccc@{}}
\toprule
\textbf{Backbone} & \textbf{GPUs} & \textbf{Wall time} & \textbf{GPU-hours} \\
\midrule
Qwen2.5-0.5B & 8  & $\sim$20h & $\sim$160 \\
Qwen2.5-1.5B & 8  & $\sim$45h & $\sim$360 \\
Qwen2.5-3B   & 24 & $\sim$40h & $\sim$960 \\
Qwen2.5-7B   & 24 & $\sim$70h & $\sim$1,680 \\
\midrule
Total & -- & -- & $\sim$3,160 \\
\bottomrule
\end{tabular*}
\end{table}

For GRPO training, we run on RTX 4090 GPUs and report the total GPU time in Table~\ref{tab:training-cost}.
In each GRPO step, SpecRL samples eight candidate annotations for the selected program, scores the group, and updates the policy. Table~\ref{tab:training-cost} aggregates the end-to-end cost of this setting across the four Qwen2.5 backbones, which takes about 3,160 GPU-hours in total.

\begin{table}[t]
\centering
\caption{Per-step time breakdown for SpecRL GRPO training.}
\label{tab:grpo-time-breakdown}
\small
\renewcommand{\arraystretch}{1.12}
\begin{tabular*}{\columnwidth}{@{\extracolsep{\fill}}lcc@{}}
\toprule
\textbf{Stage} & \textbf{Step share} & \textbf{Dafny subshare} \\
\midrule
Generation & 14.2\% & -- \\
Old log-prob & 4.8\% & -- \\
Reference log-prob & 5.0\% & -- \\
Reward + GRPO adv. & \textbf{47.8\%} & -- \\
\quad Dafny compile/resolve & -- & 18.6\% \\
\quad Dafny verification & -- & \textbf{53.5\%} \\
\quad Spectest execution & -- & 28.0\% \\
Actor update & 26.2\% & -- \\
Validation & 1.7\% & -- \\
Checkpoint/other & 0.3\% & -- \\
\bottomrule
\end{tabular*}
\end{table}

Table~\ref{tab:grpo-time-breakdown} then breaks down where a training step spends its time. Generation is the cost of sampling candidate annotations. The old-policy and reference-model log-probability rows are standard GRPO bookkeeping used to compute the policy objective. Actor update is the backpropagation and optimizer step. Validation and checkpointing are small periodic costs.
The largest block is reward scoring plus GRPO advantage computation. This is where SpecRL adds most of its training overhead.
Within reward scoring, the Dafny-backed checks are the main source of overhead. Table~\ref{tab:grpo-time-breakdown} shows that verification is the largest Dafny-side component, followed by spectest execution. Thus, spectests add a measurable cost, but verifier calls remain the dominant Dafny-side expense. 
This overhead is paid only during training; inference uses the trained model directly and does not generate spectests or filter sampled candidates.

\begin{findingbox}[Finding 4 (RQ4)]
SpecRL's extra cost comes from one-time spectest construction and GRPO reward scoring. The four SpecRL training runs cost about 3,160 GPU-hours in total, and reward scoring takes 47.8\% of GRPO step time.
\end{findingbox}

\subsection{Case Study}
\label{sec:case-study}
We use \texttt{findItem} as a case study to give an intuitive comparison of the specifications generated by different methods.
\begin{lstlisting}
method findItem(head: ObjDesc?, searchName: string)
  returns (found: bool)
  requires head != null
\end{lstlisting}
\noindent
The method searches \texttt{head.items} for a non-null element whose \texttt{name} equals \texttt{searchName}, and returns the result in \texttt{found}.
We summarize the generated suite here and give the per-test breakdown in Supplementary Appendix~A.
It contains 8 distinct spectests: 2 false-positive mutations where no matching item exists, and 6 false-negative mutations where a match exists but \texttt{found} is flipped to \texttt{false}.
Among the false negatives, 3 place the match at position~0 and 3 place it later.
We use this suite to compare the reference specification with specifications produced by ReForm and SpecRL.
The reference postcondition is:
\begin{lstlisting}
ensures found ==> exists i :: 0 <= i < |head.items|
                  && head.items[i] != null
                  && head.items[i].name == searchName
\end{lstlisting}

\noindent
This is a one-way implication. It constrains only the case \texttt{found == true} and says nothing about \texttt{found == false}. As a result, it rejects the 2 false-positive mutations but misses all 6 false-negative mutations, receiving a score of \textbf{2/8}.
ReForm adds a reverse implication, but only for the first element:
\begin{lstlisting}
ensures found ==> exists i :: 0 <= i < |head.items|
                  && head.items[i] != null
                  && head.items[i].name == searchName
ensures |head.items| > 0
        && head.items[0] != null
        && head.items[0].name == searchName ==> found
\end{lstlisting}

\noindent
This specification is already provably logically stronger than the reference specification, so the ReForm reward gives full credit and provides no further training signal. It catches the 3 false-negative mutations where the match is at position~0, but still misses the 3 where the match appears later, receiving a score of \textbf{5/8}.
SpecRL is different because its reward increases with the number of rejected spectests. The model therefore continues to receive useful signal as the specification becomes more accurate, eventually discovering the full biconditional:
\begin{lstlisting}
ensures found <==> exists i :: 0 <= i < |head.items|
                   && head.items[i] != null
                   && head.items[i].name == searchName
\end{lstlisting}

\noindent
This covers all indices in both directions, rejecting all 8 spectests and yielding a score of \textbf{8/8}.
The example illustrates the main mechanism behind SpecRL: after a reference-based reward stops changing, empirical completeness can still reward the more accurate specification.

\subsection{Threats to Validity}
\label{sec:threats}

\subsubsection{Spectest Coverage}
Our final goal is to improve specification accuracy, but there is no general oracle for deciding whether a specification exactly matches the program behavior.
We therefore use spectests as an empirical completeness signal.
Although our spectest construction (Section~\ref{sec:speccheck}) tries to target likely specification weaknesses, each suite is still finite and may miss corner cases or semantic gaps.
Thus, even if a specification rejects all generated spectests, this does not formally prove that it is fully accurate.

\subsubsection{Executability Filtering}
The current \texttt{SpecCheck} construction requires executable predicates, so programs involving non-executable specification constructs, such as ghost functions or unbounded quantifiers, are filtered out before suite construction.
The direct measure of this loss is the \textbf{SpecCheck-ready} column in the dataset statistics table in Supplementary Appendix~B.
Executable \texttt{SpecCheck} validation removes 17.0\% of Py2Dfy-Spec programs, no DafnyComp-Spec programs, and 7.1\% of DafnyBench-Spec programs.

This filter keeps the protocol well-defined but restricts the benchmark to programs where runtime spectest execution is available.
A broader backend could encode spectests as Dafny verification obligations, which would handle non-executable specifications, but it is much slower: our comparison in Supplementary Appendix~C finds an 8.01\(\times\) average slowdown over runtime checking.
Since reward scoring is already the largest part of GRPO training time and Dafny verification is much slower than spectest execution (Section~\ref{sec:cost-analysis}), using this verifier-backed backend inside RL would leave GPUs idle for extended periods while training waits for verifier calls to complete, substantially increasing training cost.

\subsubsection{Benchmark Contamination}
Because Py2Dfy, DafnyBench, and DafnyComp are public, open-source and closed-source models may have seen related programs during pretraining, so the comparison with large general-purpose models should be read cautiously.

\section{Related Work}
\emergencystretch=1em

\subsection{LLMs for Formal Verification}

Formal verification aims to prove that a program satisfies desired properties. It requires two kinds of artifacts: specifications that state the intended behavior, and auxiliary annotations, such as loop invariants, assertions, and lemmas, that help the verifier prove the implementation satisfies them.

Most LLM-based work targets the downstream proving stage, where the specification is assumed to be available and the model generates auxiliary annotations to discharge the resulting proof obligations.
Dafny is a widely used auto-active verification language that integrates specifications, code, and proofs in a single program.
Misu et al.~\cite{misu2024dafnybench} introduce DafnyBench as a benchmark for Dafny verification tasks. Laurel~\cite{mugnier2024laurel} generates helper assertions using retrieval-augmented prompting, DafnyPro~\cite{banerjee2026dafnypro} proposes an inference-time framework with diff-checking, pruning, and hint augmentation for auxiliary annotation generation, and Chakraborty et al.~\cite{chakraborty2023ranking} rank LLM-generated loop invariants to improve verification success. Baksys et al.~\cite{baksys2025minif2fdafny} translate the miniF2F mathematical benchmark into Dafny's auto-active format and evaluate LLM-based theorem proving, and Proof2Silicon~\cite{jha2025proof2silicon} targets verified Dafny code and hardware generation.
Beyond Dafny, AutoVerus~\cite{yang2025autoverus} uses iterative LLM-verifier feedback for proof generation in Verus/Rust, and Chakraborty et al.~\cite{chakraborty2025neural} study proof-oriented programming in F*.
Broader LLM-based theorem-proving systems make the same assumption: the theorem statement is given, and the model searches for a proof in systems such as Isabelle/HOL, Lean, or Coq~\cite{first2023baldur,yang2023leandojo,thakur2024copra,xin2024deepseekprover}.
SpecRL instead learns to synthesize missing Dafny specifications and auxiliary annotations, using negative tests to reward specification strength.

A related line of work asks LLMs to generate the specification itself.
Some systems generate specifications for an existing program, while others generate specifications together with code or auxiliary annotations.
AutoSpec~\cite{wen2024autospec} combines LLMs with static analysis to generate specifications for C programs, SAFE~\cite{chen2025safe} synthesizes Verus requires/ensures clauses and proofs from existing Rust code through self-evolving SFT, AlphaVerus~\cite{aggarwal2025alphaverus} translates Python programs into formally verified Verus code, and VeRuSyn~\cite{di2026verusyn} scales up SAFE's data-synthesis pipeline to 6.9 million verified Rust programs.
Starting from natural language, DafnySynthesis~\cite{misu2024dafnysynthesis} prompts large models to synthesize verified Dafny methods from Mostly Basic Python Problems (MBPP)-style tasks, including code, specifications, and auxiliary annotations.
Since its tasks are mostly short, single-method programs, we did not include it in our evaluation.
Clover~\cite{sun2024clover} simultaneously generates code, annotations, and docstrings with consistency checking, and Mirchev et al.~\cite{mirchev2024assured} co-evolve code, specifications, and tests from natural language descriptions.
Endres et al.~\cite{endres2024postconditions} study whether LLMs can translate natural-language intent into formal postconditions, and Le-Cong et al.~\cite{lecong2025formalbench} evaluate LLMs on formal specification inference across many programs.
These works show that specification synthesis is difficult even when models receive strong semantic cues.
Our stripped-program setting instead needs a scalable training signal for verifier-ready annotations on existing Dafny implementations.
The closest prior work is ReForm~\cite{yan2025reform}, which uses a reference-based binary reward that gives credit only when a candidate specification is proved logically stronger than a reference specification. We instead use a fine-grained reward based on spectest rejection, which provides an empirical completeness signal.

\subsection{Specification Testing and Validation}

Our spectest reward is based on the idea that a specification can and should be tested beyond mere verifiability.
This idea has a long history in formal methods: prior work has studied how to test, animate, debug, and validate specifications so that errors in the specification itself can be found before they affect implementation or verification~\cite{kemmerer1985testing,mccluskey1996validation,liu2016validating,konighofer2009debugging,jard1983approach}.
More recently, this line of thinking has been applied to LLM-generated specifications.
Lahiri~\cite{lahiri2024evaluating} proposes symbolic testing to assess both soundness and completeness of specifications produced by language models, a methodology later adopted by SAFE~\cite{chen2025safe}.
Our work shares the view that verification alone is insufficient, but uses execution-backed negative tests as a training reward rather than only an evaluation or filtering signal.

Dafny has some existing tools that can generate tests or assess specification accuracy, but they cannot directly replace our spectest generation pipeline.
DTest, part of the Dafny testing toolkit~\cite{fedchin2023dtest}, repurposes the verifier to generate Dafny unit-test (DUnit) cases that meet coverage goals, but it does not generate the mutated outputs needed to test whether a specification rejects impossible behaviors.
MutDafny~\cite{amaral2025mutdafny} mutates the implementation and checks whether the specification detects the injected fault, whereas our setting keeps the given method implementation fixed and tests the accuracy of specifications against that implementation.

Symbolic specification synthesis has also been studied in domain-specific language (DSL) settings.
Spyro~\cite{park2023spyro} synthesizes strongest properties for a user-provided semantic query and DSL, and LOUD~\cite{park2025loud} extends this idea to existential queries and both over- and under-approximating specifications.
Their guarantees are tied to the supplied query and DSL; applying them to general Dafny programs would require modeling Dafny's rich verification semantics, and they do not generate auxiliary annotations such as loop invariants, intermediate assertions, or lemma calls.
SpecRL instead directly generates verifier-ready Dafny annotations from existing programs, using verifier feedback and negative tests to improve specification strength.

\subsection{RL for Code Generation}

Reinforcement learning has been applied to code-related tasks with verifier or execution feedback as reward signals. CodeRL~\cite{le2022coderl} trains code generation models with unit test feedback via actor-critic RL. StepCoder~\cite{dou2024stepcoder} decomposes code generation into a curriculum of subtasks with compiler feedback. RLEF~\cite{gehring2025rlef} teaches models to use execution feedback for multi-turn code synthesis. RLCoder~\cite{wang2025rlcoder} uses RL to train a retriever for repository-level code completion. SWE-RL~\cite{wei2025swerl} trains on open-source software evolution data for software engineering tasks. CURE~\cite{wang2025cure} co-evolves code generators and unit testers via RL, with tests providing feedback for code correctness. These rewards mainly optimize generated code behavior, whereas SpecRL scores verified specifications by how well they reject implementation-impossible behaviors.

\section{Conclusion}

We presented SpecRL, a reinforcement learning framework for synthesizing specifications and auxiliary annotations for existing Dafny programs.
SpecRL addresses a central limitation of verifier-only feedback: a candidate specification can verify while still admitting behaviors the implementation cannot produce.
To distinguish such weak specifications from more accurate ones, SpecRL builds spectests offline and uses their rejection rate as an empirical completeness signal during training.
On the out-of-distribution DafnyComp-Spec benchmark, SpecRL improves both empirical completeness and verification success over SFT and RL with a reference-based binary reward across four model scales.

\section*{Acknowledgments}
OpenAI Codex\footnote{\url{https://openai.com/codex/}} was used to assist with the visual presentation of Figure~\ref{fig:specrl-overall} and with language polishing in the manuscript. The authors reviewed and approved all AI-assisted edits and take full responsibility for the technical content, experimental results, and conclusions.

\appendices


\section{findItem Case Study Details}
\label{sec:appendix-finditem}

This appendix provides the detailed qualitative evidence behind Section~4.3 of the main paper: the full per-test breakdown for \texttt{findItem}.

\noindent\textbf{Method under test.}
\begin{lstlisting}
method findItem(head: ObjDesc?, searchName: string)
  returns (found: bool)
  requires head != null
{
  found := false;
  var i := 0;
  while i < |head.items|
  {
    if head.items[i] != null
       && head.items[i].name == searchName {
      found := true;
      return;
    }
    i := i + 1;
  }
}
\end{lstlisting}

\noindent\textbf{Spectests and per-test results.}
Table~\ref{tab:case-combined} lists the 8 non-duplicate spectests and shows whether each specification level rejects them.
Each spectest first runs the method on a concrete input, then mutates \texttt{found} to the opposite value and checks whether the specification rejects the mutated result.
A \ding{55} means the specification correctly rejected the mutation (desirable), and a \ding{51} means the mutation escaped (undesirable).

The 8 spectests fall into three groups.
Tests~1--2 are false positives: no matching item exists, so the observed result is \texttt{false}, but the spectest mutates it to \texttt{true}.
These cases are already handled by the reference postcondition \texttt{found ==> exists ...}, so all three specifications reject them.
Tests~3--5 are false negatives where the matching item is at position~0.
The observed result is \texttt{true}, but the spectest mutates it to \texttt{false}; the reference specification misses them because it says nothing when \texttt{found == false}, while ReForm and SpecRL reject them through a reverse implication that covers the first element.
Tests~6--8 are false negatives where the matching item appears after position~0.
These cases expose the main limitation of ReForm: its reverse constraint only mentions \texttt{head.items[0]}, so it still accepts the mutated \texttt{false} result, whereas SpecRL rejects all of them because its biconditional ranges over all positions.

This grouping directly explains the rejection rates in the last row.
The reference specification rejects only the 2 false positives, yielding 2/8.
ReForm rejects those same 2 spectests plus the 3 false negatives with a match at position~0, yielding 5/8.
SpecRL rejects all 8 tests, yielding 8/8.

\begin{table*}[h!]
\centering
\caption{Spectests for \texttt{findItem} and per-level rejection results. Each test calls the method, mutates \texttt{found} to the opposite value, and checks whether the specification rejects the mutation. \ding{55}\,=\,rejected (good), \ding{51}\,=\,escaped (bad).}
\label{tab:case-combined}
\small
\renewcommand{\arraystretch}{1.15}
\begin{tabular}{clllccc}
\toprule
\textbf{\#} & \textbf{Input (\texttt{head.items}, \texttt{searchName})} & \textbf{Observed} & \textbf{Mutated} & \textbf{Reference} & \textbf{ReForm} & \textbf{SpecRL} \\
\midrule
1 & \texttt{[null, null]}, \texttt{"any"}                      & \texttt{false} & \texttt{true}  & \ding{55} & \ding{55} & \ding{55} \\
2 & \texttt{[id("x"), id("y")]}, \texttt{"z"}                  & \texttt{false} & \texttt{true}  & \ding{55} & \ding{55} & \ding{55} \\
3 & \texttt{[id("a")]}, \texttt{"a"}                           & \texttt{true}  & \texttt{false} & \ding{51} & \ding{55} & \ding{55} \\
4 & \texttt{[id("a"), null]}, \texttt{"a"}                     & \texttt{true}  & \texttt{false} & \ding{51} & \ding{55} & \ding{55} \\
5 & \texttt{[id("x"), id("y")]}, \texttt{"x"}                  & \texttt{true}  & \texttt{false} & \ding{51} & \ding{55} & \ding{55} \\
6 & \texttt{[null, id("target")]}, \texttt{"target"}            & \texttt{true}  & \texttt{false} & \ding{51} & \ding{51} & \ding{55} \\
7 & \texttt{[id("a"), id("b")]}, \texttt{"b"}                  & \texttt{true}  & \texttt{false} & \ding{51} & \ding{51} & \ding{55} \\
8 & \texttt{[null, null, id("c")]}, \texttt{"c"}               & \texttt{true}  & \texttt{false} & \ding{51} & \ding{51} & \ding{55} \\
\midrule
  & \multicolumn{3}{l}{\textbf{Rejection rate}} & 2/8 & 5/8 & 8/8 \\
\bottomrule
\end{tabular}
\end{table*}


\section{Benchmark Details}
\label{sec:appendix-datasets}

\subsection{Benchmark Descriptions}
\label{sec:appendix-benchmarks}

Our evaluation uses three specification-focused benchmarks, each augmented from an existing Dafny dataset and covering a different difficulty level.

\noindent\textbf{Py2Dfy-Spec.}
Py2Dfy-Spec is derived from Py2Dfy, a collection of Python programs automatically translated into Dafny, with both stripped code and fully annotated reference code provided as paired data. The upstream Python-to-Dafny translation was performed prior to this work; our pipeline only performs filtering, format conversion, and spectest construction. Its lower retention rate is not mainly caused by verification failure. Instead, many translated files contain class wrappers, nested logic, and nontrivial dependencies across declarations, which makes it harder to identify a suitable top-level target method and to assemble self-contained, executable spectests. The processed benchmark provides 4,389 programs for RL training and 274 programs for validation-based checkpoint selection.

\noindent\textbf{DafnyComp-Spec.}
DafnyComp-Spec is derived from DafnyComp, a competition-level dataset translated from programming contest problems into Dafny. The retained programs are much longer and more specification-heavy than those in Py2Dfy-Spec, making DafnyComp-Spec a more challenging out-of-distribution benchmark.

\noindent\textbf{DafnyBench-Spec.}
DafnyBench-Spec is derived from DafnyBench~\cite{misu2024dafnybench}. The source dataset contains 606 verified Dafny programs, but 469 are single-method programs, making the full retained benchmark relatively easy for our setting. Spectest construction retains 428 programs. We therefore use the longest quartile of the retained programs, yielding 107 programs with 1,059 spectests for the additional evaluation.

\noindent\textbf{Release timing and potential contamination.}
Py2Dfy and DafnyComp were released through ReForm in July 2025~\cite{yan2025reform}. As with other public code benchmarks, pretraining contamination is difficult to avoid completely in practice once the data have been publicly available.

\subsection{Dataset Statistics}
\label{sec:appendix-stats}

We apply the same validity check and three cumulative construction filters to all three source datasets; each column in Table~\ref{tab:dataset-stats} counts only programs retained by the preceding column.
\textbf{Valid} counts programs whose reference code passes \texttt{dafny verify}.
\textbf{Target-ready} counts valid programs for which the current construction pipeline can identify a suitable executable target method, assemble its required source context into a self-contained test file, and extract a \texttt{SpecCheck} predicate.
Programs excluded by this filter generally do not expose a suitable target under the current harness-construction constraints; they therefore never become optimization or evaluation targets, rather than representing cases where SpecRL optimization fails. As detailed below, one Py2Dfy-Spec program is an exception because a late re-verification check disagreed with its preprocessing result.
\textbf{SpecCheck-ready} counts Target-ready programs whose generated executable \texttt{SpecCheck} predicate compiles and passes Dafny validation.
\textbf{Suite-ready} counts SpecCheck-ready programs for which execution-backed input-output pairs and negative spectests are successfully constructed and executed. As a positive sanity check, the reference \texttt{SpecCheck} predicate must accept every input-output pair observed from actual program execution.
\textbf{Used} counts the Suite-ready programs selected for the experiments; for DafnyBench-Spec, this is the longest quartile rather than the full retained suite.
The last four columns characterize the Used programs: \textbf{Avg LOC} is the average number of non-empty lines in the reference program, \textbf{Avg methods} is the average number of methods per program, \textbf{Avg spec} is the average number of \texttt{requires}, \texttt{ensures}, and \texttt{invariant} clauses per program, and \textbf{Avg Spectests} is the average number of generated spectests per program.

\begin{table*}[t]
\centering
\caption{Dataset construction statistics. Counts are cumulative from Valid through Used; average columns are computed over Used programs.}
\label{tab:dataset-stats}
\small
\setlength{\tabcolsep}{2.2pt}
\renewcommand{\arraystretch}{1.05}
\begin{tabular*}{\textwidth}{@{\extracolsep{\fill}}l rrrrr rrrr}
\toprule
\textbf{Dataset} & \textbf{Valid} & \textbf{\shortstack{Target-\\ready}} & \textbf{\shortstack{SpecCheck-\\ready}} & \textbf{\shortstack{Suite-\\ready}} & \textbf{Used} & \textbf{Avg LOC} & \textbf{Avg methods} & \textbf{Avg spec} & \textbf{Avg Spectests} \\
\midrule
Py2Dfy-Spec      & 15{,}906 & 6{,}761 & 5{,}615 & 4{,}663 & 4{,}663 & 61.1  & 2.2 & 12.9 & 11.6 \\
DafnyComp-Spec   & 262      & 256     & 256     & 232     & 232     & 227.4 & 5.7 & 43.1 & 12.6 \\
DafnyBench-Spec  & 606      & 480     & 446     & 428     & 107     & 77.4  & 3.0 & 14.5 & 9.9 \\
\bottomrule
\end{tabular*}
\end{table*}

Several observations emerge from the statistics:
\begin{itemize}
    \item \textbf{Py2Dfy-Spec} is by far the largest benchmark, but it also undergoes the strongest reduction through the filtering pipeline.
    \item \textbf{DafnyComp-Spec} is the most demanding benchmark: its programs are much longer, contain substantially more specification clauses and loop invariants, and therefore require richer reasoning than Py2Dfy-Spec.
    \item \textbf{DafnyBench-Spec} contains many short single-method programs, so we use the longest quartile of its retained programs for the additional evaluation.
\end{itemize}

By default, the spectest construction pipeline asks for five concrete inputs per target method and three output mutations per input, then removes duplicate input-output pairs and generated test files that fail to compile or execute. After this filtering, Py2Dfy-Spec contains an average of 11.64 spectests per retained program, with a median of 13. DafnyComp-Spec contains an average of 12.56 spectests per retained program, with a median of 12. The full retained DafnyBench-Spec suite contains an average of 9.52 spectests per program, with a median of 10; the 107-program evaluation slice contains 1,059 spectests, or 9.90 spectests per program on average. These counts explain the rounded ``Avg Spectests'' column in Table~\ref{tab:dataset-stats}.

The filtering losses are construction losses, not model-performance filtering. Across all three datasets, Target-ready filtering primarily excludes files for which the pipeline cannot identify a suitable executable target or assemble its required declarations into a self-contained test file. SpecCheck-ready filtering excludes generated predicates that do not compile or pass Dafny validation, while Suite-ready filtering excludes programs without usable observed input-output pairs, valid negative spectests, or successful positive sanity checks.

For Py2Dfy-Spec, Target-ready filtering removes 9,145 valid programs: 7,737 fall outside the current top-level-method test-file construction, one is a late re-verification mismatch in the preprocessing logs, and 1,407 fail target-method or \texttt{SpecCheck} extraction. This category primarily covers files whose methods are nested inside classes or non-unwrapped modules, files whose only top-level methods are \texttt{Main} or test drivers, and files whose dependencies cannot be packaged with a standalone target. SpecCheck-ready filtering then removes 1,146 programs, or 17.0\% of the 6,761 Target-ready programs; most failures involve unresolved identifiers or dependencies (897 cases), followed by parse/syntax failures (97), ghost constructs in executable contexts (23), type/resolution errors (23), non-compilable quantifiers (20), and other validation failures (86). Suite-ready filtering removes another 952 programs: 920 lack valid observed pairs or negative spectests, 23 fail positive sanity checks, and 9 fail because of API errors. For DafnyComp-Spec, Target-ready filtering removes 6 programs, SpecCheck-ready filtering removes none, and Suite-ready filtering removes 24, almost all because no valid observed pairs or negative spectests remain. For DafnyBench-Spec, the three filters remove 126, 34, and 18 programs, respectively; we then use the longest 107 retained programs for the additional evaluation.


\section{Spectest Checking Backend}
\label{sec:appendix-speccheck}

SpecRL checks spectests by compiling a runtime-checkable \texttt{SpecCheck} predicate and executing it with \texttt{dafny test}.
An alternative is to encode each spectest as a symbolic verification obligation, as in verifier-in-the-loop specification evaluation~\cite{lahiri2024evaluating,chen2025safe}.
Runtime checking is narrower because the predicate must be executable, but it is deterministic for verified programs\footnote{Dafny's verifier proves termination of every method, function, and loop by default. The \texttt{SpecCheck} predicate is a non-recursive predicate syntactically extracted from the method's specification clauses, so its evaluation also always terminates.} and much faster in our measurements.
Symbolic verification is broader because it can handle non-executable specifications, but it is slower and can be unstable under fixed time limits.
We therefore use runtime checking for online RL and present symbolic verification as a compatible backend.

We illustrate both backends using the \texttt{Max} example from Section~3.2 of the main paper.
Both use the same weak specification from Figure~1 of the main paper and the same four mutated outputs; only the checking mechanism differs.

\subsection{Approach A: Runtime Checking (Ours)}
\label{sec:appendix-runtime}

The runtime backend assembles the file below and executes it with \texttt{dafny test --no-verify}.
The generated \texttt{SpecCheck} predicate is executable code: each spectest calls the method, mutates the output, and checks \texttt{expect !SpecCheck(...)}.
For executable predicates, spectest checking reduces to ordinary execution on concrete inputs.

\begin{lstlisting}
method Max(a: array<nat>) returns (m: int)
  requires a.Length > 0
  ensures forall k :: 0 <= k < a.Length
            ==> m >= a[k]
{
  if a.Length == 0 { return -1; }
  var i := 0;
  m := a[0];
  while i < a.Length
    invariant 0 <= i <= a.Length
    invariant forall k :: 0 <= k < i ==> m >= a[k]
  { if a[i] >= m { m := a[i]; } i := i + 1; }
}

// SpecCheck predicate (compiled from requires/ensures clauses)
predicate SpecCheck(a: array<nat>, m: int)
{
  a.Length > 0 ==>
    forall k :: 0 <= k < a.Length ==> m >= a[k]
}

// --- Negative tests (mutated outputs) ---

method {:test} Neg_IO1_1() {
  var a := new nat[] [3, 1, 4, 1, 5];
  var m := Max(a);
  m := 14;  // sum of elements
  expect !SpecCheck(a, m);  // FAILED: 14 >= all, accepted
}

method {:test} Neg_IO1_2() {
  var a := new nat[] [3, 1, 4, 1, 5];
  var m := Max(a);
  m := 100;  // arbitrary large value
  expect !SpecCheck(a, m);  // FAILED: 100 >= all, accepted
}

method {:test} Neg_IO1_3() {
  var a := new nat[] [3, 1, 4, 1, 5];
  var m := Max(a);
  m := 3;  // non-maximum element
  expect !SpecCheck(a, m);  // PASSED: 3 < 5
}

method {:test} Neg_IO1_4() {
  var a := new nat[] [];
  var m := Max(a);
  m := 99;  // empty-array behavior
  expect !SpecCheck(a, m);  // FAILED: precondition false, accepted
}
\end{lstlisting}

\noindent
\textbf{Result}: 1/4 tests pass.
The weak specification accepts $m{=}14$ and $m{=}100$ because both satisfy $m \geq a[k]$ for all $k$, but without a membership check such as \texttt{ensures m in a[..]}, these spurious values cannot be rejected.
It also accepts the empty-array mutation because the precondition \texttt{a.Length > 0} is false, making \texttt{SpecCheck} vacuously true.

\subsection{Approach B: Symbolic Verification}
\label{sec:appendix-symbolic}

This file is checked with \texttt{dafny verify}, where each spectest is encoded as a verification obligation that the candidate specification rejects the mutated output.

\begin{lstlisting}
method Max(a: array<nat>) returns (m: int)
  requires a.Length > 0
  ensures forall k :: 0 <= k < a.Length
            ==> m >= a[k]
{
  if a.Length == 0 { return -1; }
  var i := 0;
  m := a[0];
  while i < a.Length
    invariant 0 <= i <= a.Length
    invariant forall k :: 0 <= k < i ==> m >= a[k]
  { if a[i] >= m { m := a[i]; } i := i + 1; }
}

// --- Spectests (verified symbolically) ---

method Neg_IO1_1(a: array<nat>) returns (m: int)
  ensures !(
    a.Length > 0 ==>
      forall k :: 0 <= k < a.Length ==> m >= a[k]
  )  // VERIFY FAILED: 14 >= all, spec accepts
{
  assume a.Length == 5;
  assume a[0]==3 && a[1]==1 && a[2]==4 && a[3]==1 && a[4]==5;
  m := 14;
}

method Neg_IO1_2(a: array<nat>) returns (m: int)
  ensures !(
    a.Length > 0 ==>
      forall k :: 0 <= k < a.Length ==> m >= a[k]
  )  // VERIFY FAILED: 100 >= all, spec accepts
{
  assume a.Length == 5;
  assume a[0]==3 && a[1]==1 && a[2]==4 && a[3]==1 && a[4]==5;
  m := 100;
}

method Neg_IO1_3(a: array<nat>) returns (m: int)
  ensures !(
    a.Length > 0 ==>
      forall k :: 0 <= k < a.Length ==> m >= a[k]
  )  // VERIFY OK: 3 < a[4]=5, negation holds
{
  assume a.Length == 5;
  assume a[0]==3 && a[1]==1 && a[2]==4 && a[3]==1 && a[4]==5;
  m := 3;
}

method Neg_IO1_4(a: array<nat>) returns (m: int)
  ensures !(
    a.Length > 0 ==>
      forall k :: 0 <= k < a.Length ==> m >= a[k]
  )  // VERIFY FAILED: precondition false, spec accepts
{
  assume a.Length == 0;
  m := 99;
}
\end{lstlisting}

\noindent
\textbf{Result}: 1/4 assertions verified.
The outcome is identical to Approach~A, but each spectest becomes a separate SMT-backed verification obligation.
The symbolic verification backend does not require the specification to be compilable, making it applicable to ghost-dependent or quantifier-heavy specifications that Approach~A cannot handle.
Moreover, the number of reported errors is not stable enough for fine-grained reward design: a single verification obligation may produce multiple error messages, so raw error counts do not reliably reflect how weak the specification is.

\subsection{Timing Comparison on Representative Programs}
\label{sec:appendix-timing}

Table~\ref{tab:speccheck-timing} compares runtime checking and symbolic verification on 10 randomly selected programs, with five sampled from each benchmark.
The Py2Dfy-Spec programs are \texttt{728 Count}, \texttt{257 ApplyPagination}, \texttt{849 GetContent}, \texttt{160 ConsumeNumbers}, and \texttt{980 RunGameSession}; the DafnyComp-Spec programs are \texttt{94 countLargestGroup\_1399}, \texttt{177 countLargestGroup\_1399}, \texttt{188 main\_3node\_2}, \texttt{194 main\_4node\_4}, and \texttt{205 main\_5node\_8}.

\begin{table}[h!]
\centering
\caption{Average spectest-checking times and symbolic-verification slowdown on 10 randomly selected programs.}
\label{tab:speccheck-timing}
\small
\setlength{\tabcolsep}{2.2pt}
\renewcommand{\arraystretch}{1.12}
\resizebox{\columnwidth}{!}{%
\begin{tabular}{@{}lrrrr@{}}
\toprule
\textbf{Dataset} & \textbf{Programs} & \textbf{Runtime (ms)} & \textbf{Symbolic (ms)} & \textbf{Slowdown} \\
\midrule
Py2Dfy-Spec & 5 & 978.25 & 1,921.07 & 1.96\(\times\) \\
DafnyComp-Spec & 5 & 945.65 & 13,487.11 & 14.26\(\times\) \\
\midrule
\textbf{Overall} & \textbf{10} & \textbf{961.95} & \textbf{7,704.09} & \textbf{8.01\(\times\)} \\
\bottomrule
\end{tabular}%
}
\end{table}

On Py2Dfy-Spec, symbolic verification takes 1,921.07\,ms per program on average, compared with 978.25\,ms for runtime checking, a 1.96\(\times\) slowdown.
The gap is much larger on DafnyComp-Spec: symbolic verification takes 13,487.11\,ms on average and is 14.26\(\times\) slower than runtime checking.
Across all 10 programs, symbolic verification is 8.01\(\times\) slower overall, supporting our use of runtime checking in the online RL loop.

\bibliographystyle{IEEEtran}
\bibliography{bibtex}

\end{document}